\documentclass[fleqn]{article}
\usepackage{hyperref}
\usepackage{authblk}

\usepackage{graphicx}
\usepackage{amsmath}
\usepackage{amsfonts}
\usepackage{amssymb}

\newcommand{\su}{\uparrow}
\newcommand{\sd}{\downarrow}

\title{Spin-constrained Hartree-Fock and the generator coordinate method for the 2-site Hubbard model. }

\author[1]{Stijn De Baerdemacker\thanks{Corresponding Author: stijn.debaerdemacker@unb.ca}}
\author[1]{Amir Ayati}
\author[2]{Hugh G.\ A.\ Burton}
\author[3]{Xeno De Vriendt}
\author[3]{Patrick Bultinck}
\author[3]{Guillaume Acke}
\affil[1]{Department of Chemistry, University of New Brunswick, 30 Dineen Drive, E3B~5A1, Fredericton Canada}
\affil[2]{Physical and Theoretical Chemistry Laboratory, Department of Chemistry, University of Oxford, South Parks Road, OX1~3QZ, Oxford, United Kingdom}
\affil[3]{Department of Chemistry, Ghent University, Krijgslaan 281-S3, 9000, Ghent, Belgium}

\begin{document}
\maketitle

\begin{abstract}
We present a mathematical analysis of the spin-constrained Hartree-Fock solutions (CHF) of the 2-site Hubbard model. The analysis sheds light on the spin symmetry breaking process around the Coulson-Fischer point.  CHF states are useful as input states for the Generator Coordinate Method (GCM) in which CHF states can be used as a basis for multiconfigurational calculations in the Hill-Wheeler equations.   The spin degree of freedom provides the generator coordinate in the GCM related to static electron correlation. \newline
\end{abstract}

\section{The generator coordinate and\\ constrained Hartree-Fock method}

\paragraph{Static electron correlation} For the vast majority of situations in chemistry, the single-reference mean-field picture is an excellent starting point for all computational purposes.  In these situations, coupled cluster theory with respect to the Hartree-Fock mean field state \cite{bartlett:2007}, or Kohn-Sham density-functional theory built on the Kohn-Sham Slater determinant \cite{parr:1989} are largely sufficient to capture the weak quantum correlations in the system.  However, whenever quantum correlations are strong, the single-reference picture breaks down, even qualitatively, and there is a need to move to multi-reference approaches \cite{helgaker:2000}.  For instance, in bond-breaking processes, the system is characterized by resonances between pairs of electrons in quasi-degenerate bonding and antibonding orbitals \cite{cooper:2002}, which is generally referred to as strong \emph{static} correlation.  Suitable methods to accurately capture strong static electron correlation include multi-configuration self-consistent field theory \cite{helgaker:2000}, density matrix renormalization group \cite{baiardi:2020} or geminal theory \cite{tecmer:2022}.  

\paragraph{Spin symmetry} An interesting alternative approach to strong static correlation is based on spin-symmetry breaking and restoration \cite{jimenez-hoyos:2012}.  When relaxing the spin-singlet restriction in the Hartree-Fock mean-field variational optimization process, allowing both the spin-up and spin-down sectors to optimize into different molecular orbitals, it is typically observed that the exact potential energy surface is much better reproduced beyond the bond-breaking distance compared to the spin-restricted case in which molecular orbitals for the spin-up and spin-down sector are identical \cite{coulson:1949}.  These calculations are commonly referred to as unrestricted Hartree-Fock (UHF) and restricted Hartree-Fock calculations (RHF) respectively.   An important consequence of the spin restriction relaxation in UHF is that the associated UHF Slater determinant is no longer an eigenstate of the total spin operator, in contrast to the RHF Slater determinant, leading to unphysical so-called spin-contamination in the wave function \cite{szabo:1996}.  In essence, we have arrived at what L\"owdin identified as the ``symmetry dilemma'' in Hartree-Fock theory, in which one has to choose between either a wave function with good symmetry quantum numbers but incorrect energy characteristics (RHF), or an improper wave function with a qualitatively correct bond-dissociation profile (UHF)\cite{lykos:1963}.  For completeness, the spin projection quantum number is still preserved in UHF, which is no longer the case if we allow the spin-up and spin-down sector to mix up completely, as in generalized Hartree-Fock (GHF) \cite{lowdin:1992}.

\paragraph{Spin projection and holomorphic Hartree Fock} An elegant way out of the symmetry dilemma is by means of symmetry projection, in which the wave function components with good spin quantum number are projected out of the UHF Slater determinant using a suitable projection operator \cite{jimenez-hoyos:2012,ring:2004}.  This is a purely mathematical procedure built upon an averaged spin-symmetry $SU(2)$ group action on the broken symmetry UHF state, ignoring the energetic properties of the UHF state.  However, methods like spin-projected Hartree--Fock only work when the symmetry breaks spontaneously at the mean-field level, creating a discontinuitiy in the energy at the Coulson-Fischer point. The non-Hermitian holomorphic Hartree--Fock approach identifies complex-analytic extensions to solutions that disappear, creating a continuous basis for subsequent nonorthogonal configuration interaction (NOCI) calculations \cite{hiscock:2014}.  The common denominator between the two mentioned approaches is that it transforms a mean-field single-reference Slater determinant into a multireference wave function, necessary for the adequate description of static correlation in bond-breaking processes. 

\paragraph{Generator Coordinate Method} Both the spin-symmetry projection and HHF as a NOCI method can be regarded as special cases of the generator coordinate method (GCM) \cite{ring:2004}.  In its most general formulation, the GCM ansatz $|\Psi\rangle$ is a wavefunction expansion along a continuous manifold of basis states \cite{griffin:1957}.  
\begin{equation}
|\Psi\rangle=\int da f(a) |\psi(a)\rangle,
\end{equation}
in which $a$ is the, possibly multivariate, generator coordinate, $|\psi(a)\rangle$ is a manifold of many-body states, and $f(a)$ are the continuous expansion functions.  For simplicity, we choose the functions $f(a)$ to be real valued.  The key to a successful GCM implementation is the choice of the generator coordinate that generates the manifold of states.  The GCM, also known as integral transform method in this context, has been explored previously for the electronic structure of atomic systems.   There, the wave function ans\"atze of one- and two-electron systems are expanded over a continuum manifold of Slater-type \cite{somojai:1968} and Hylleraas-type \cite{thakkar:1977} wave functions respectively, giving the generator coordinate a physical interpretation as an effective charge.  The GCM has also been explored for molecular structure applications beyond the Born-Oppenheimer Approximation, in which the molecular wavefunctions have been expanded along nuclear coordinates, inducing a coupling between the nuclear and electronic degrees of freedom \cite{lathouwers:1978}.  Arguably the most successful application of the GCM can be found in nuclear structure physics \cite{bender:2003}, where the generator coordinate also gets a geometric interpretation as it characterizes the collective multipole deformation modes of the nucleus \cite{reinhard:1987}.  For deformed nuclei, these multipole deformation modes are associated with spontaneous symmetry breaking, as the nucleus has non-zero multipole moments away from the ideal spherical droplet model.    

The advantage of choosing a physically motivated generator coordinate $a$ is that one can expect \textit{a priori} that the expansion functions $f(a)$ will be localized around physically meaningful values of $a$, with the width of the function interpreted as fluctuations in $a$.  Nevertheless, the generic approach to compute the functions $f(a)$ is via the variational optimization of the energy functional
\begin{equation}
E[f] = \frac{\langle\Psi|\hat{H}|\Psi\rangle}{\langle\Psi|\Psi\rangle}, 
\end{equation}
with the stability condition 
\begin{equation}
\frac{\delta E[f]}{\delta f(a)}=0,
\end{equation}
leading to the so-called Hill-Wheeler (HW) equations \cite{ring:2004}
\begin{equation}
\int db \langle\psi(a)|\hat{H}|\psi(b)\rangle f(b) = E\int db \langle\psi(a)|\psi(b)\rangle f(b).\label{hillwheeler}
\end{equation}
This is a generalized eigenvalue problem with a non-orthogonal overlap kernel $O(a,b)=\langle\psi(a)|\psi(b)\rangle$ and Hamiltonian kernel $H(a,b)=\langle\psi(a)|\hat{H}|\psi(b)\rangle$ over a continuous manifold.  
\paragraph{Constrained Hartree-Fock} From a practitioner's point of view, it is key to choose a manifold of computationally tractable input states $|\psi(a)\rangle$, and discretize the integrals by sampling the generator coordinate $a$ sufficiently dense around their physically meaningful values.  Both objectives can be obtained by means of the constrained Hartree-Fock (CHF) method, in which the constraint is imposed on the relevant physical observable $\hat{A}$ associated with the generator coordinate $a$
\begin{equation}
\langle \hat{A}\rangle = a. 
\end{equation}
First, CHF is a Hartree-Fock mean-field method, yielding a Slater determinant, for which computationally facile expressions for the overlap and Hamiltonian kernel exist.  Second, it is fairly straightforward to impose constraints by virtue of the Hellmann-Feynman theorem, which is applicable for all variational approaches, hence also the Hartree-Fock approach.  For this, we define the Lagrangian
\begin{equation}
\hat{\mathcal{L}}= \hat{H}-\mu(\hat{A}-a), 
\end{equation}
which immediately encodes the imposed constraint when varying with respect to the Lagrange multiplier $\mu$
\begin{equation}
\frac{\partial\langle\hat{\mathcal{L}}\rangle}{\partial\mu}=(\langle \hat{A}\rangle - a )\equiv 0,
\end{equation}
in the variationally obtained mean-field Slater determinant wavefunction 
\begin{equation}
|\psi(a)\rangle=|\textrm{HF}(a)\rangle. 
\end{equation}
The idea of imposing constraints in Hartree-Fock wavefunctions is common practice in nuclear structure physics, and has been proposed in quantum chemistry as well, for instance in the context of fixing erroneous dipole moments in UHF wavefunctions \cite{mukherji:1963}, or uncovering phase transitions in magnetic fields \cite{lemmens:2022}.  

The modern implementation of GCM is to identify the observable associated with symmetries that break spontaneously as the generator coordinate and construct a manifold, or discretized set, of CHF states to solve the HW equations as a non-orthogonal configuration interaction (NOCI) problem.  In nuclear structure physics, the broken symmetry is associated with the multipole deformation of the nucleus.  In electronic structure theory, the idea of feeding non-orthogonal Hartree-Fock states into a NOCI framework has been proposed in the past, however not associated with broken symmetries\cite{thom:2009}.  In a recent \cite{devriendt:2022} and forthcoming \cite{ayati:2023} publication, we propose respectively the spin projection $M$ and total spin observable $S$ as the physically relevant generator coordinate for strongly correlated molecular systems.  

In the present chapter, we dive deeper into the GCM process, by mathematically analyzing a 2-site Hubbard model, a minimal toy model of the H${}_2$ molecule in a minimal basis set.  The simplicity of the model allows us to obtain closed expressions for the CHF, while retaining all the important and non-trivial physical attributes and characteristics of the  H${}_2$ dimer.  In addition, it provides a clean playground to investigate the NOCI process for solving the HW equations.  

In the next section \ref{section:hubbard}, we introduce the 2-site Hubbard model with its exact eigenstates and spin properties.  In the following section \ref{section:chf}, we analyze the CHF solutions of the model and discuss the symmetry breaking features.  In section \ref{section:gcm}, we report the results of a minimal GCM calculation, and present conclusions in section \ref{section:conclusions}.
\section{The Hubbard model}\label{section:hubbard}
\begin{figure}[!htb]
\begin{center}
\includegraphics{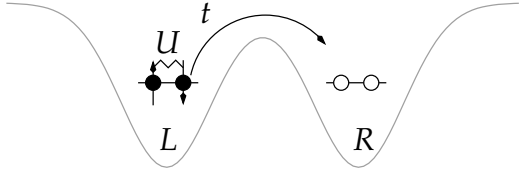}
\caption{Cartoon picture of the 2-site Hubbard model.  The orbitals localized around the atomic centers have been replaced by simplified lattice sites, labeled by $L$ and $R$.  The inter-site hopping $t$ is represented by an arrow, whereas the Coulomb repulsion $U$ is denoted by a wiggly line.  The state depicted is $|1\rangle$ in Fock space (\ref{2site:fockspace}), and will generate state $|3\rangle$ after the hopping of the spin-down electron took place. }\label{figure:cartoon}
\end{center}
\end{figure}

\paragraph{The model} The Hubbard model is a popular minimal model describing the electronic structure in theoretical condensed matter systems \cite{fradkin:2013}.  The premise of the model is that each atom in the lattice provides one or multiple orbitals, or lattice sites, which the electrons can dynamically occupy.  The appeal of the model is in the abstraction of the chemical specificities of the interactions, which simplifies the 1-body kinetic energy and electron-nuclei interactions to an electron-hopping interaction that respects the geometric graph structure of the lattice, and reduces the 2-body electron-electron interaction to just a local on-site Coulomb repulsion term.  The Hamiltonian in second quantization becomes 
%%becomes 
\begin{equation}
\hat{H}_{\textrm{Hub}}=-\sum_{ij}t_{ij}\sum_{\sigma=\su,\sd}(\hat{a}_{i\sigma}^\dag \hat{a}_{j\sigma}+\hat{a}_{j\sigma}^\dag \hat{a}_{i\sigma}) + U\sum_{i}\hat{n}_{i\su}\hat{n}_{i\sd},\label{hubbard:hamiltonian}
\end{equation}
with the operators $\hat{a}^\dag_{i\sigma}$ and $\hat{a}_{i\sigma}$ creating or annihilating an electron with spin $\sigma$ in orbital $i$ respectively, $\hat{n}_{i\sigma}=\hat{a}^\dag_{i\sigma}\hat{a}_{i\sigma}$ the number operator counting the number of spin-$\sigma$ electrons in orbital $i$, $t$ the hopping matrix and $U$ the on-site Coulomb repulsion energy.  The hopping matrix $t$ encodes the graph structure of the lattice, and is typically restricted to local nearest-neighbour interactions.  Both the hopping amplitudes $t_{ij}$ and Coulomb repulsion energy $U$ are chosen positive.  The creation/annihilation operators obey Fermi-Dirac statistics, imposed by the anti-commutation relations
\begin{equation}
\{\hat{a}^\dag_{i\sigma},\hat{a}_{j\tau}\}=\delta_{ij}\delta_{\sigma\tau},\quad\{\hat{a}_{i\sigma},\hat{a}_{j\tau}\}=0,\quad\{\hat{a}^\dag_{i\sigma},\hat{a}^\dag_{j\tau}\}=0.
\end{equation}
The physics described by the Hubbard Hamiltonian is a competition between the delocalizing 1-body hopping interaction $t$ and the local on-site 2-body Coulomb repulsion $U$, leading to strong quantum correlations in the model.  Historically, in order to avoid the high multi-reference character of the model, the on-site 2-body interactions have been effectively absorbed in the 1-body terms in what is known as the tight-binding approximation \cite{slater:1954}, leading to a free model described by single-reference states of delocalized molecular orbitals.  From a chemist's point of view, the tight-binding approximation of finite-size molecules is also known as the H\"uckel model, the celebrated model underpinning many of the heuristic rules in, e.g., organic chemistry \cite{zimmerman:1975}.  As a result, the original Hubbard Hamiltonian (\ref{hubbard:hamiltonian}) for finite-size molecular systems can be regarded as a correlated H\"uckel model in which the 2-body interactions have not been effectively renormalized in the 1-body interactions, providing an ideal test ground for many multi-reference approaches.  
\paragraph{Spin properties} Each individual lattice site in the Hubbard model carries an $SU(2)$ Schwinger representation 
\begin{equation}
\hat{S}_i^\dag = \hat{a}_{i\su}^\dag\hat{a}_{i\sd},\quad \hat{S}_i= (\hat{S}_i^\dag)^\dag=\hat{a}_{i\sd}^\dag\hat{a}_{i\su},\quad \hat{S}_i^0=\frac{1}{2}(\hat{n_{i\su}}-\hat{n}_{i\sd}),
\end{equation}
with canonical $SU(2)$ commutation relations
\begin{equation}
[\hat{S}_i^{0},\hat{S}_j^\dag]=\hat{S}_i^\dag\delta_{ij},\quad[\hat{S}_i^{0},\hat{S}_j]=-\hat{S}_i\delta_{ij},\quad[\hat{S}_i,\hat{S}_j]=2\hat{S}^0_{i}\delta_{ij},
\end{equation}
and spin magnitude operators
\begin{equation}
\hat{S}^2_i=\tfrac{1}{2}[\hat{S}_i^\dag\hat{S}_i+\hat{S}_i\hat{S}_i^\dag]+(\hat{S}_i^0)^2=\tfrac{3}{4}(\hat{n}_{i\su}-\hat{n}_{i\sd})^2,
\end{equation}
in which the idempotency of the number operators has been used $\hat{n}_{i\sigma}^2=\hat{n}_{i\sigma}$.  The only allowed representations are either the fundamental spin-$\frac{1}{2}$ doublet representation, or the two trivial representations consisting of the vacuum and completely filled state.  The total spin algebra of the full system is then obtained by a simple recoupling of all lattice spins. 
\paragraph{2-site Hubbard model} The simplest finite-size Hubbard Hamiltonian is the 2-site Hubbard dimer, which provides a minimal toy model  for the H${}_2$ dimer.  The Hamiltonian (\ref{hubbard:hamiltonian}) is simplified to
\begin{equation}
\hat{H}_{2} =  -t\sum_{\sigma=\su,\sd}(\hat{a}_{L\sigma}^\dag \hat{a}_{R\sigma}+\hat{a}_{R\sigma}^\dag \hat{a}_{L\sigma}) + U(\hat{n}_{L\su}\hat{n}_{L\sd}+\hat{n}_{R\su}\hat{n}_{R\sd})\label{2site:hamiltonian}
\end{equation}
in which $L$ and $R$ label the Left and Right (hydrogen) site of the system respectively and $t$ describes the hopping strength between the two sites.  The strength of the hopping parameter $t$ is related to the proximity of the two sites; how farther the two sites, the weaker the hopping strength, with $t\rightarrow0$ in the limit of infinite distance.  As the energy units are arbitrary, we will employ the description $(t,U)=(1-\xi,\xi)$ with $\xi\in[0,1]$ to cover the full range of interactions with a single parameter $\xi$.  A cartoon picture of the 2-site Hubbard model is given in Figure \ref{figure:cartoon}

\begin{figure}[!htb]
\begin{center}
\includegraphics{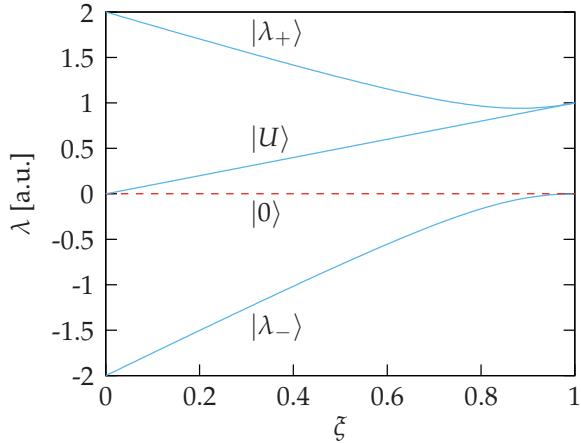}
\caption{Energy levels of the 2-site Hubbard model (\ref{2site:hamiltonian}) as a function of the parameter $\xi$ with $(t,U)=(1-\xi,\xi)$ in arbitrary units. The states are labeled by their energy eigenvalues.  The full lines denote $S=0$ singlet states, whereas the dashed line represents the 3-fold degenerate $S=1$ triplet state ($M=-1,0,+1$).}\label{figure:fci}
\end{center}
\end{figure}

We will only consider the ``neutral'' 2-site Hubbard model, which would amount to a total of 2 electrons.  The full Fock space of the 2-particle sector is spanned by the states
\begin{align}
&|1\rangle = a_{L\su}^\dag a_{L\sd}^\dag|\theta\rangle,\quad |2\rangle = a_{R\su}^\dag a_{R\sd}^\dag|\theta\rangle,\quad
|3\rangle = a_{L\su}^\dag a_{R\sd}^\dag|\theta\rangle,\quad |4\rangle = a_{R\su}^\dag a_{L\sd}^\dag|\theta\rangle,\notag\\
&|5\rangle = a_{L\su}^\dag a_{R\su}^\dag|\theta\rangle,\quad |6\rangle = a_{L\sd}^\dag a_{R\sd}^\dag|\theta\rangle.\label{2site:fockspace}
\end{align}
with $|\theta\rangle$ the empty particle vacuum.  The first four states are $M=0$ eigenstates of the total spin projection operator, whereas $|5\rangle$ and $|6\rangle$ are $M=+1$ and $M=-1$ eigenstates respectively.  The total spin operators are defined by
\begin{equation}
\hat{S}^\dag =\hat{S}^\dag_L+\hat{S}^\dag_R,\quad \hat{S} =\hat{S}_L+\hat{S}_R,\quad \hat{S}^0 =\hat{S}^0_L+\hat{S}^0_R\label{2site:totalspin},
\end{equation}
and the magnitude by 
\begin{equation}
\hat{S}^2=\tfrac{1}{2}[\hat{S}^\dag\hat{S}+\hat{S}\hat{S}^\dag]+(\hat{S}^0)^2.\label{2site:spin2} 
\end{equation}
As the Hamiltonian (\ref{2site:hamiltonian}) commutes with the total spin projection $\hat{S}^0$, Fock space breaks down into the three spin projection components $M=\{0,\pm1\}$.  We can focus on the $M=0$ subspace spanned by the states $\{|1\rangle,|2\rangle,|3\rangle,|4\rangle\}$.  The four exact eigenstates $|\lambda\rangle$ of the Hamiltonian (\ref{2site:hamiltonian}), labeled by their eigenvalues $\lambda$ are
\begin{itemize}
\item The constant-energy $\lambda=0$ state  
\begin{equation}
|\lambda=0\rangle=\tfrac{1}{\sqrt{2}}(|3\rangle-|4\rangle)=\tfrac{1}{\sqrt{2}}(a_{L\su}^\dag a_{R\sd}^\dag + a_{L\sd}^\dag a_{R\su}^\dag)|\theta\rangle
\end{equation}
is the $M=0$ state of the $S=1$ triplet, complemented by $|5\rangle$ and $|6\rangle$.
\item The $\lambda=U$ doubly-occupied $S=0$ closed singlet state 
\begin{equation}
|\lambda=U\rangle=\tfrac{1}{\sqrt{2}}(|1\rangle-|2\rangle)=\tfrac{1}{\sqrt{2}}(a_{L\su}^\dag a_{L\sd}^\dag - a_{R\su}^\dag a_{R\sd}^\dag)|\theta\rangle
\end{equation}
\item The $S=0$ de/constructive $\lambda_{\pm}=\frac{1}{2}U\pm\sqrt{(\frac{U}{2})^2+(2t)^2}$ resonance states
\begin{equation}
|\lambda_\pm\rangle=\cos\theta_\pm\tfrac{1}{\sqrt{2}}(|1\rangle+|2\rangle)+\sin\theta_{\pm}\tfrac{1}{\sqrt{2}}(|3\rangle+|4\rangle),
\end{equation}
with 
\begin{equation}
\tan\theta_{\pm}=\tfrac{U}{4t}\mp\sqrt{\left(\tfrac{U}{4t}\right)^2+1}.
\end{equation}  
Note that $\tan\theta_+\tan\theta_-=-1$, such that $\theta_+-\theta_-=\frac{\pi}{2}$, pointing out that $|\lambda_{\pm}\rangle$ are indeed mutually orthogonal.  The $|\lambda_-\rangle$ state is the ground state.  
\end{itemize}
The exact energy levels of the Hamiltonian are shown in Figure plotted \ref{figure:fci} as a function of $\xi\in[0,1]$, which tunes the interaction from the pure hopping ($\xi=0\rightarrow U=0$) to suppressed hopping $(\xi=1\rightarrow t=0)$ regime. 
Because of the dimensionless parameter $\xi$, all energy values are given in arbitrary units [a.u.].  The energy spectrum and eigenstates are fairly well understood over the full range of the parameter $\xi$, partially due to their solutions in analytic form.  Whereas the $\xi\rightarrow0$ limit is dominated by a single reference state characterised by maximally delocalized electrons over both sites, the $\xi\rightarrow1$ has a perfect degeneracy between the singlet ground state $|\lambda_-\rangle$ and triplet state $|\lambda=0\rangle$, consistent with a picture of strongly localized electrons on both individual sites.   
\section{Restricted, Unrestricted\\ and Constrained Hartree-Fock}\label{section:chf}
\begin{figure}[!htb]
\begin{center}
\includegraphics{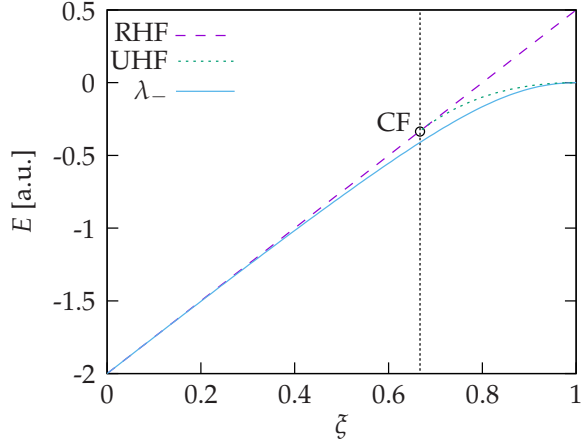}
\caption{RHF (\ref{rhf:energy}) and UHF (\ref{uhf:energy}) energies in arbitrary units for the 2-site hubbard system, compared to the energy of the exact ground state  $|\lambda_-\rangle$. The Coulson Fischer point (CF) is denoted by means of a dotted line at $\xi=\frac{2}{3}$. }\label{figure:uhf}
\end{center}
\end{figure}

\begin{figure}[!htb]
\begin{center}
\includegraphics{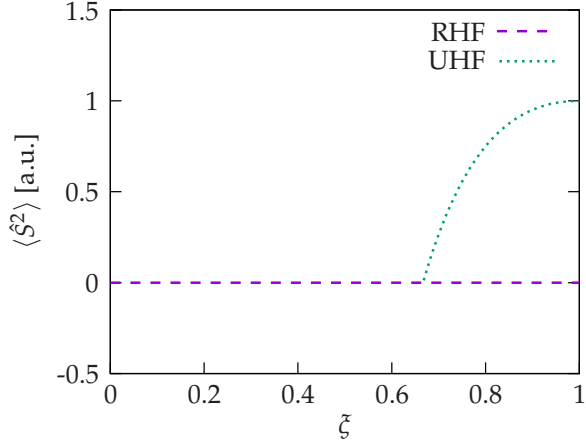}
\caption{Expectation values of the spin operator $\hat{S}^2$ in arbitrary units for the RHF and UHF states of the 2-site hubbard model as a function of $\xi$.}\label{figure:spin}
\end{center}
\end{figure}

\begin{figure}[!htb]
\begin{center}
\includegraphics{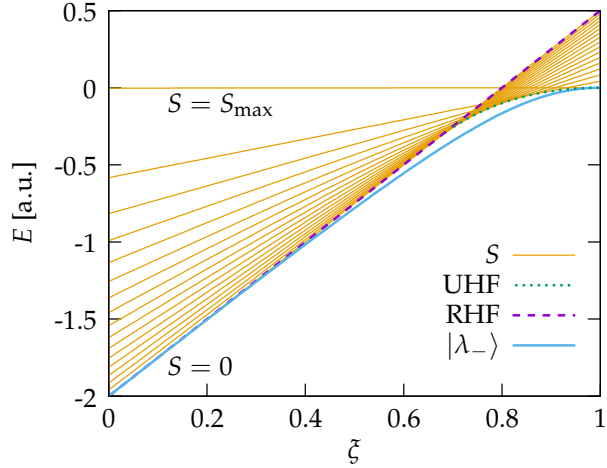}
\caption{CHF energies (\ref{chf:energy}) for different values of the constrained spin value $S$.  RHF, UHF and exact ground state energies are given for easy reference.  Curves with increasing intercept at $\xi=0$ correspond to increasing $S$ from $S=0$ to $S=S_{\textrm{max}}$.  The $S=0$ curve coincides with the RHF prediction, whereas the $S=S_{\textrm{max}}$ has constant $E=0$.}\label{figure:chf:energy}
\end{center}
\end{figure}

\begin{figure}[!htb]
\begin{center}
\includegraphics{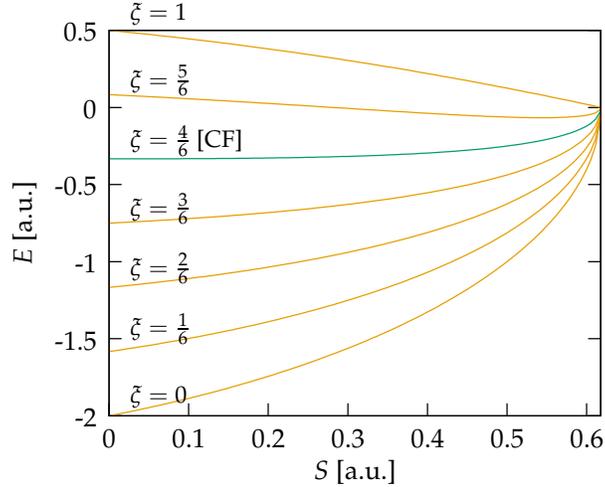}
\caption{CHF energies (\ref{chf:energy}) as a function of $S$ for different values of $\xi\in\{0,\frac{1}{6},\dots,1\}$.  The Coulson-Fischer point $\xi=\frac{4}{6}$ is labeled with [CF].}\label{figure:chf:spin}
\end{center}
\end{figure}

\paragraph{(Un)restricted Hartree-Fock} Before embarking into the spin-constrained CHF, it is instructive to revisit the general structure of the UHF and RHF solutions of the 2-site Hubbard model (\ref{2site:hamiltonian}).  The most general spin-unrestricted Slater determinant state is parametrized by
\begin{align}
|\textrm{SD}\rangle &= (\cos\alpha \hat{a}_{L\su}^\dag + \sin\alpha\hat{a}_{R\su}^\dag)(\cos\beta \hat{a}_{L\sd}^\dag+\sin\beta\hat{a}_{R\sd}^\dag)|\theta\rangle \label{chf:sd}\\
&=\cos\alpha\cos\beta|1\rangle+\sin\alpha\sin\beta|2\rangle+\cos\alpha\sin\beta|3\rangle+\sin\alpha\cos\beta|4\rangle\label{chf:fockspaceexpansion}
\end{align}
in which the parameters $\alpha$ and $\beta$ parametrize the molecular single-particle states of the spin-up and spin-down electrons respectively, and the second line, eq.\ (\ref{chf:fockspaceexpansion}), is an expansion in Fock space (\ref{2site:fockspace}) for all practical purposes.  We note that setting $\alpha=\beta$ in (\ref{chf:sd}) spin-restricts the Slater determinant.  The variational energy functional of the state is
\begin{align}
E[\alpha,\beta]&=\langle\textrm{SD}|\hat{H}_2|\textrm{SD}\rangle \\
&=-t[\sin2\alpha+\sin2\beta]+\tfrac{1}{2}U[1+\cos2\alpha\cos2\beta]\label{sd:energy}
\end{align}
leading to the stationarity conditions
\begin{align}
\frac{\partial E}{\partial\alpha}&=-2t\cos2\alpha-U\cos2\beta\sin2\alpha = 0\\
\frac{\partial E}{\partial\beta} &=-2t\cos2\beta -U\cos2\alpha\sin2\beta = 0
\end{align}
Among the many solutions, the following are of particular interest:
\begin{itemize}
\item The trivial solution $\alpha=\beta=\frac{\pi}{4}$ ($\cos2\alpha=\cos2\beta=0$) is the RHF solution
\begin{equation}
|\textrm{RHF}\rangle = \tfrac{1}{2}(\hat{a}_{L\su}^\dag + \hat{a}_{R\su}^\dag)(\hat{a}_{L\sd}^\dag+\hat{a}_{R\sd}^\dag)|\theta\rangle.
\end{equation}
The energy (\ref{sd:energy}) associated with this state is 
\begin{equation}
E[\tfrac{\pi}{4},\tfrac{\pi}{4}]=-2t+\tfrac{1}{2}U,\label{rhf:energy}
\end{equation}
which is exact in the $U=0$ ($\xi=0$) pure hopping limit. 
\item The solution associated with $\sin2\alpha=\sin2\beta=\frac{2t}{U}$ does not necessarily refer to spin-restricted solutions, since $\alpha$ and $\beta$ are not necessarily equal.  Indeed, the stationarity conditions say that $\cos2\alpha+\cos2\beta=0$, leading to the solution $\alpha+\beta=\frac{\pi}{2}$.  This is the UHF solution.  Note that this solution is only possible whenever
\begin{equation}
\frac{2t}{U}\le1,
\end{equation}
because of the constraints imposed by the trigonometric nature of this solution $\sin2\alpha=\sin2\beta=\frac{2t}{U}$.  The point for which $\frac{2t}{U}=1$ corresponds to the Coulson-Fischer point, as the UHF solution becomes equivalent to the RHF solution. For values of $U$ larger than $2t$, it is energetically favorable to (artificially) break the symmetry, leading to the UHF energy being lower than the RHF state.  The energy (\ref{uhf:energy}) associated with this solution is
\begin{equation}
E[\alpha_{\textrm{UHF}},\tfrac{\pi}{2}-\alpha_{\textrm{UHF}}]=-\frac{1}{2}\frac{(2t)^2}{U},\label{uhf:energy}
\end{equation}
with $\alpha_{\textrm{UHF}}=\frac{1}{2}\arccos\sqrt{1-(\frac{2t}{U})^2}$.
\end{itemize}
The RHF (\ref{rhf:energy}) and UHF (\ref{uhf:energy}) energies are presented in Figure \ref{figure:uhf} in comparison with the exact energy of the ground state $|\lambda_-\rangle$ as a function of $\xi$.
The RHF energy is a linear function of $\xi$, whereas the UHF energy beyond the Coulson-Fischer point follows the exact energy curve slightly better with a quadratic behaviour.  In the parametrisation ($t,U$)=($1-\xi,\xi$), the Coulson-Fischer point happens at $\xi=\frac{2}{3}$.
\paragraph{Spin symmetry} It is straightforward to verify that the Coulson-Fischer point indeed corresponds to a spontaneous symmetry breaking by computing the total spin expectation value and spin contamination.  The expectation value of the total spin operator (\ref{2site:spin2}) for the general Slater determinant (\ref{chf:sd}) is
\begin{equation}
\langle\textrm{SD}|\hat{S}^2|\textrm{SD}\rangle = \sin^2(\alpha-\beta),
\end{equation}
which leads to
\begin{align}
\langle\textrm{RHF}|\hat{S}^2|\textrm{RHF}\rangle =& 0\label{rhf:spin}\\
\langle\textrm{UHF}|\hat{S}^2|\textrm{UHF}\rangle =& 1-\left(\tfrac{2t}{U}\right)^2\label{uhf:spin}
\end{align}
for the RHF and UHF state respectively.  Both functions are presented in Figure \ref{figure:spin} as a function of $\xi$. 
The spin of the RHF state is trivially zero.  However, as soon as the Coulson-Fischer point is reached at $\xi=\frac{2}{3}$ or $\frac{2t}{U}=1$, the UHF solution implies a non-zero spin expectation value, indicative of a broken symmetry.  It is remarkable to note that the spin value for the UHF solution (\ref{uhf:spin}) in the $\xi\rightarrow1$ ($t\rightarrow0$) limit converges to 1 instead of 2, leading to $S=\frac{\sqrt{5}-1}{2}\neq 1$. The reason for this is the incapability of the UHF state (\ref{chf:sd}) to describe a pure $S=1$ triplet state, such as for the $|\lambda=0\rangle$ state in the exact spectrum.  The maximal spin attainable by the UHF state will be $S_{\textrm{max}}=\frac{\sqrt{5}-1}{2}$.   This observation can be generalized to $N$-electron systems, where fully spin-polarized states with $S=\frac{N}{2}$ should be treated by generalized Hartree-Fock (GHF).
\paragraph{Constrained Hartree-Fock} We are now positioned to analyse the constrained Hartree-Fock (CHF) states of the model.  In essence, CHF is equivalent to UHF with an additional constraint on the spin observable, which is imposed by means of a Lagrange multiplier.  The CHF Hamiltonian becomes
\begin{equation}
\hat{\mathcal{H}}_2=\hat{H}_2-\mu[\hat{S}^2-S(S+1)]\label{chf:hamiltonian}
\end{equation}
with $\hat{H}_2$ the original 2-site Hubbard Hamiltonian (\ref{2site:hamiltonian}), the total spin operator given by eq.\ (\ref{2site:spin2}), and $\mu$ the Lagrange multiplier.  The value $S$ is externally imposed by the user, and can vary between $S=0$ and $S=S_{\textrm{max}}$.  The expectation value of (\ref{chf:hamiltonian}) in the Slater determinant state (\ref{chf:sd}) is
\begin{align}
E[\alpha,\beta;\mu] =& -t[\sin2\alpha+\sin2\beta]+\tfrac{1}{2}U[1+\cos2\alpha\cos2\beta]\\
&\qquad-\mu[\sin^2(\alpha-\beta)-S(S+1)]
\end{align}
The stationarity conditions are slightly augmented by means of the spin constraint
\begin{align}
\frac{\partial E}{\partial\alpha}&=-2t\cos2\alpha-U\cos2\beta\sin2\alpha -\mu\sin(2\alpha-2\beta) = 0\\
\frac{\partial E}{\partial\beta} &=-2t\cos2\beta -U\cos2\alpha\sin2\beta +\mu\sin(2\alpha-2\beta) = 0\\
\frac{\partial E}{\partial\mu} & = -\sin^2(\alpha-\beta)+S(S+1) = 0. 
\end{align}
The spin constraint introduces a new coupling between the $\alpha$ and $\beta$ variables, however the solutions can still be obtained in closed mathematical form.  For this, consider the coordinate transformation
\begin{equation}
\sigma=\alpha-\beta,\quad\tau=\alpha+\beta,
\end{equation} 
leading to the elegant energy expression
\begin{align}
E[\sigma,\tau;\mu]=&-2t\sin\tau\cos\sigma +\tfrac{1}{2}U[1+\cos^2\sigma-\sin^2\tau]\\
&\quad-\mu[\sin^2\sigma - S(S+1)],
\end{align}
and the simplified stationarity equations
\begin{align}
\frac{\partial E}{\partial \tau} & = -2t\cos\tau\cos\sigma-U\sin\tau\cos\tau = 0,\label{chf:dEdtau}\\
\frac{\partial E}{\partial \sigma} & =+2t\sin\tau\sin\sigma-U\cos\sigma\sin\sigma-\mu\sin2\sigma=0,\label{chf:dEdsigma}\\
\frac{\partial E}{\partial \mu}& =-\sin^2\sigma+S(S+1)=0.\label{chf:dEdlambda}
\end{align}
First, we can solve for the variable $\sigma$ via (\ref{chf:dEdlambda})
\begin{equation}
\sin\sigma=\sqrt{S(S+1)},\label{chf:sigma}
\end{equation}
in which we have chosen the positive solution.  We will assume everywhere that both $\alpha$ and $\beta$ will be contained in the domain $[0,\frac{\pi}{2}]$ and that $\alpha\ge\beta$.  Other solutions can all be mapped to this domain.  The value for $\tau$ can now be deduced from (\ref{chf:dEdtau}), which has two solutions
\begin{equation}
\cos\tau=0,\qquad\textrm{or}\qquad 2t\cos\sigma+U\sin\tau=0.
\end{equation}
Only the first solution corresponds to a local minimum, which is familiar from the UHF solution
\begin{equation}
\tau=\alpha+\beta=\tfrac{\pi}{2}.
\end{equation}
Finally, the value for the Lagrange multiplier can be easily obtained from eq.\ (\ref{chf:dEdsigma})
\begin{equation}
\mu=\tfrac{t}{\sqrt{1-S(S+1)}}-\tfrac{1}{2}U.\label{chf:lambda}
\end{equation}
Bringing everything together leads to the energy expectation value 
\begin{equation}
E(t,U,S)=-2t\sqrt{1-S(S+1)}+\tfrac{1}{2}U[1-S(S+1)],\label{chf:energy}
\end{equation}
as a function of $t$, $U$ and the imposed spin $S$.  The CHF is presented in Figure \ref{figure:chf:energy} as a function of $\xi$ for different constrained spin values, and in Figure (\ref{figure:chf:spin}) as a function of $S$ for different values of $\xi$. 
There are several aspects to note:

The first aspect is that all CHF energy curves (\ref{chf:energy}) are straight lines as a function of $\xi$ for fixed values of $S$.  The $S=0$ CHF curve is identical to the RHF energy (\ref{rhf:energy})
\begin{equation}
E(t,U,S=0)=-2t+\tfrac{1}{2}U.
\end{equation}
A closely related observation is that each $S\neq0$ CHF energy  is tangential to the UHF energy curve.  More precisely, each CHF line with specific $S$ touches the UHF curve at the $\xi$ (or $(t,U)$) value for which the UHF spin (\ref{uhf:spin}) is exactly $S$
\begin{equation}
E(t,U,S_{\textrm{UHF}})=-\tfrac{1}{2}\tfrac{(2t)^2}{U},
\end{equation}
with $S_{\textrm{UHF}}$ extracted from (\ref{uhf:spin}).  In a related aspect, the spin value $S$ for which the CHF energy is minimal
\begin{equation}
\frac{\partial E}{\partial S} = (2S+1)\left[\tfrac{t}{\sqrt{1-S(S+1)}}-\tfrac{1}{2}U\right]=0,
\end{equation}
equally leads to the spin expectation value of UHF
\begin{equation}
S(S+1) = 1-\left(\tfrac{2t}{U}\right)^2,
\end{equation}
and UHF energy (\ref{uhf:energy}) respectively. Note that this solution is only physically acceptable whenever $\frac{2t}{U}\le 1$, matching with the Coulson-Fischer criterion.

A second aspect concerns a physical interpretation of the Lagrange multiplier $\mu$.  From eq,\ (\ref{chf:lambda}), one can infer the necessary value of $\mu$ to impose a certain spin $S$.  Inverting the relation gives rise to 
\begin{equation}
S(S+1)=1-\left(\tfrac{2t}{2\mu+U}\right)^2=1-\left(\tfrac{2t}{U_{\textrm{eff}}}\right)^2,
\end{equation}
pointing out that the Lagrange multiplier $\mu$ provides us with a means to tune the effective on-site Coulomb energy $U_{\textrm{eff}}=2\mu+U$, such that the spin symmetry can be explicitly broken in the UHF towards the desired spin value.

A final aspect is related to the spontaneous symmetry breaking around the Coulson-Fischer point $\frac{2t}{U}=1$ ($\xi=\frac{2}{3}$).  For this, we present the CHF energies as a function of $S$ for different values of $\xi$ in Figure \ref{figure:chf:spin}.
We observe an evolution of the CHF energy as the parameter $\xi$ is increased from 0 to 1.  For all values of $\xi$ below the Coulson-Fischer point ($\xi\le\frac{2}{3}$), the CHF has a variational minimum at $S=0$.  In contrast, above the Coulson-Fischer point ($\xi>\frac{2}{3}$), the variational minimum is found at $S\neq0$.  This is corroborated by the evaluating the slope of the CHF energy at the $S=0$ intercept
\begin{equation}
\left.\frac{\partial E}{\partial S}\right|_{S=0}=t-\tfrac{1}{2}U
\end{equation}
The slope is strictly positive for $t>\frac{1}{2}U$, and changes sign at exactly the Coulson-Fischer point, again an indicator of the spin symmetry breaking phase transition.
\section{Generator Coordinate Method }\label{section:gcm}
\begin{figure}[!htb]
\begin{center}
\includegraphics{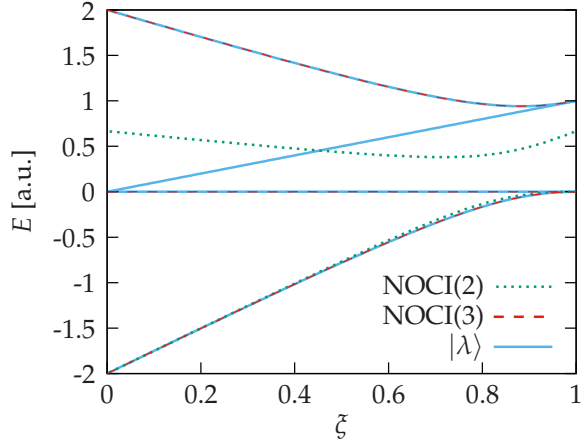}
\caption{NOCI($n$) GCM calculations for the 2-site hubbard model with an equidistant grid in $S$ for $n=2$ (dotted lines) and $n=3$ (dashed lines).  The exact eigenvalues are presented as reference (solid lines).  All units are arbitrary [a.u.]. }\label{figure:noci}
\end{center}
\end{figure}

\paragraph{Spin GCM} The idea behind the GCM is to use a manifold of spin-constrained CHF states
\begin{equation}
|\Psi\rangle = \int_0^{S_{\textrm{max}}}dS f(S) |\textrm{CHF}(S)\rangle,
\end{equation}
to solve the Hill-Wheeler equations (\ref{hillwheeler})
\begin{equation}
\int_0^{S_{\textrm{max}}}dS_1  \langle S_2|\hat{H}_2|S_1\rangle f(S_1)=E\int_0^{S_{\textrm{max}}}dS_1\langle S_2|S_1\rangle f(S_1)
\end{equation}
in which the $|S\rangle$ states are a shorthand notation for the constrained Hartree-Fock states (\ref{chf:sd}) to a certain spin $S$. The overlap kernel and Hamiltonian kernel are given by 
\begin{align}
\langle S_2|S_1\rangle &= \tfrac{1}{2}[1+\cos(\sigma_1-\sigma_2)]\\
\langle S_2|\hat{H}_2|S_2\rangle &= -t(\cos\sigma_1+\cos\sigma_2)+\tfrac{1}{2}U\cos\sigma_1\cos\sigma_2, 
\end{align}
with $\sigma_1$ and $\sigma_2$ the solution of the spin constraint equation (\ref{chf:sigma}).  Note that we regain the CHF energies (\ref{chf:energy}) for the selected CHF states on the diagonal when $\sigma_1=\sigma_2$.  
\paragraph{Non-orthogonal configuration interaction} For all practical purposes, the Hill-Wheeler equations need to be discretized, in a finite set of basis states
\begin{equation}
\sum_{S_1} \langle S_2|\hat{H}_2|S_1\rangle f(S_1)=E\sum_{S_1}\langle S_2|S_1\rangle f(S_1),
\end{equation} 
leading to the secular equation of non-orthogonal configuration interaction (NOCI), as the CHF states with different $S$ are non-orthogonal in general.  There are multiple ways to choose the set of discrete $S$ values, and optimizing for the best grid will strongly depend on the system of interest \cite{devriendt:2022,bonche:1990}.  This study falls outside the scope of the present analysis, so we envision a simple rule here.  We take $n$ equidistant points in the $S\in[0,S_{\textrm{max}}]$ interval, and use the CHF states at those $n$ spin values as the grid.  The results for $n=2$ and $n=3$ are presented in Figure \ref{figure:noci} as NOCI(2) and NOCI(3).  
It is interesting to note that NOCI(2) captures the ground state $|\lambda_-\rangle$ fairly accurately, whereas NOCI(3) also provides a reasonable description of the $|\lambda=0\rangle$ and $|\lambda_+\rangle$ excited states.  This is noteworthy as the $|\lambda=0\rangle$ state is a $S=1$ triplet configuration.  Going to NOCI(4) allows us to span the full Hilbert space of the exact problem and is equivalent to the exact calculation.  
\section{Conclusions}\label{section:conclusions}
\paragraph{Constrained Hartree Fock} We have analyzed the mathematical features of a spin constrained Hartree-Fock solution of the 2-site Hubbard model.  The closed form of the solutions follow from the symmetry and simplicity in the model.  The spontaneous spin symmetry breaking has been characterized explicitly, and a physically meaningful interpretation has been given to the Lagrange multiplier in terms of an effective on-site repulsion. 

\paragraph{Generator Coordinate Method} Subsequently, we have employed the CHF states into a discretized version of the Hill-Wheeler equations, and reproduced the exact ground state of the model at both sides of the Coulson-Fischer point, showing that the generator coordinate method should be able to capture strong static correlation effects that are outside single-reference mean field methods.  The present study can serve as a simple map to guide more sophisticated GCM calculations for more intricate systems.  
\section*{Acknowledgment}
We acknowledge conversations with Alex Thom on the holomorphic Hartree-Fock method, which has been the inspiration for the spin GCM.  SDB and AA acknowledge the Canada Research Chair program, the CFI, NSERC and NBIF for financial support. HGAB thanks New College, Oxford for funding through the Astor Junior Research Fellowship.  Parts of this research were funded by FWO research project G031820N. 
%%
%\bibliography{sdb-aiqc}

\end{document}